\RequirePackage[columnwise]{lineno}
\documentclass[aps,prl,twocolumn,floatfix,final,superscriptaddress,citeautoscript,longbibliography]{revtex4}
\usepackage{amssymb}
\usepackage{graphicx}
\usepackage{amsmath}
\usepackage{braket}
\usepackage[colorlinks=true,linkcolor=blue,citecolor=blue,urlcolor=blue]{hyperref}
\usepackage{xcolor}
\usepackage{lineno}
\begin{document}

%%%%%% title
\title{Interrelated Thermalization and Quantum Criticality in a Lattice Gauge Simulator}

%%%%%% authors
\author{Han-Yi Wang}
\thanks{H.-Y.W.~, W.-Y.Z.~and Z.-Y.Y.~contributed equally to this work.}
\affiliation{Hefei National Research Center for Physical Sciences at the Microscale and School of Physics, University of Science and Technology of China, Hefei 230026, China}

\author{Wei-Yong Zhang}
\thanks{H.-Y.W.~, W.-Y.Z.~and Z.-Y.Y.~contributed equally to this work.}
\affiliation{Hefei National Research Center for Physical Sciences at the Microscale and School of Physics, University of Science and Technology of China, Hefei 230026, China}

\author{Zhi-Yuan Yao}
\thanks{H.-Y.W.~, W.-Y.Z.~and Z.-Y.Y.~contributed equally to this work.}
\affiliation{Institute for Advanced Study, Tsinghua University, Beijing 100084, China}

\author{Ying Liu}
\affiliation{Hefei National Research Center for Physical Sciences at the Microscale and School of Physics, University of Science and Technology of China, Hefei 230026, China}

\author{Zi-Hang Zhu}
\affiliation{Hefei National Research Center for Physical Sciences at the Microscale and School of Physics, University of Science and Technology of China, Hefei 230026, China}

\author{Yong-Guang Zheng}
\affiliation{Hefei National Research Center for Physical Sciences at the Microscale and School of Physics, University of Science and Technology of China, Hefei 230026, China}

\author{Xuan-Kai Wang}
\affiliation{Hefei National Research Center for Physical Sciences at the Microscale and School of Physics, University of Science and Technology of China, Hefei 230026, China}

\author{Hui Zhai}
\affiliation{Institute for Advanced Study, Tsinghua University, Beijing 100084, China}
\affiliation{Hefei National Laboratory, University of Science and Technology of China, Hefei 230088, China}

\author{Zhen-Sheng Yuan}
\affiliation{Hefei National Research Center for Physical Sciences at the Microscale and School of Physics, University of Science and Technology of China, Hefei 230026, China}
\affiliation{Hefei National Laboratory, University of Science and Technology of China, Hefei 230088, China}
\affiliation{CAS Center for Excellence in Quantum Information and Quantum Physics, University of Science and Technology of China, Hefei 230026, China}

\author{Jian-Wei Pan}
\affiliation{Hefei National Research Center for Physical Sciences at the Microscale and School of Physics, University of Science and Technology of China, Hefei 230026, China}
\affiliation{Hefei National Laboratory, University of Science and Technology of China, Hefei 230088, China}
\affiliation{CAS Center for Excellence in Quantum Information and Quantum Physics, University of Science and Technology of China, Hefei 230026, China}

%%%%%% Abstract part.
\begin{abstract}
Gauge theory and thermalization are both foundations of physics and nowadays are both topics of essential importance for modern quantum science and technology \cite{Wiese:2013uq,Tagliacozzo:2013oa,Zohar:2015qs,Dalmonte:2016lg,Banuls:2020sl,Banuls:2020ro,Aidelsburger:2022ca,Nandkishore:2015mb,Bloch2019,Ueda:2020qe}. Simulating lattice gauge theories (LGTs) realized recently with ultracold atoms provides a unique opportunity for carrying out a correlated study of gauge theory and thermalization in the same setting \cite{Yang:2020oo,Zhou:2021td}. Theoretical studies have shown that an Ising quantum phase transition exists in this implemented LGT \cite{Sachdev:2002mi,Fendley:2004cd,Rico:2014tn,Surace:2020lg,Yao:2022qm}, and quantum thermalization can also signal this phase transition \cite{Yao:2022qm}. Nevertheless, it remains an experimental challenge to accurately determine the critical point and controllably explore the thermalization dynamics in the quantum critical regime due to the lack of techniques for locally manipulating and detecting matter and gauge fields. Here, we report an experimental investigation of the quantum criticality in the LGT from both equilibrium and non-equilibrium thermalization perspectives by equipping the single-site addressing and atom-number-resolved detection into our LGT simulator. We accurately determine the quantum critical point agreed with the predicted value \cite{Sachdev:2002mi,Fendley:2004cd,Rico:2014tn}. We prepare a $\ket{\mathbb{Z}_2}$ state deterministically and study its thermalization dynamics across the critical point, leading to the observation that this $\ket{\mathbb{Z}_2}$ state thermalizes only in the critical regime \cite{Yao:2022qm}. This result manifests the interplay between quantum many-body scars, quantum criticality, and symmetry breaking.
\end{abstract}

% make title
\date{\today}
\maketitle

Since quantum gauge theories are computationally intractable in the non-perturbative regime, the idea of formulating gauge theories on discretized space-time lattices led to LGTs, and the developments of LGTs enable numerical simulation of gauge theories with various classical algorithms in the past decades \cite{Rothe:Book}. Recently, a new trend is realizing LGTs in quantum simulators with ultracold atoms or trapped ions \cite{Martinez:2016rt,Kokail:2019sv,Schweizer:2019fa,Mil:2020as,Yang:2020oo,Zhou:2021td}. By taking the quantum advantage, these quantum simulation platforms offer the promise of a better understanding of LGTs than classical simulations.

One potential advantage of quantum simulation for studying LGTs lies in the non-equilibrium dynamics, such as quantum thermalization. In ultracold atom systems, the essential parameters in the gauge theory are tunable, allowing accessing different phases and the quantum criticality in between. Quantum criticality refers to a qualitative change of the ground state and the universal behavior of low-energy physics. On the contrary, thermalization dynamics usually involves highly excited states. Remarkably, it has been recently proposed that the thermalization dynamics of the $\ket{\mathbb{Z}_2}$ state can also signal the quantum critical regime in the U(1) LGT \cite{Yao:2022qm}. However, it is challenging to study these physics due to the lack of techniques for manipulating and detecting local matter and gauge fields in the previous experiments.

In this work, we report on an experimental study of both equilibrium and thermalization dynamics in the quantum critical regime of the U(1) LGT. We integrate the techniques of single-site manipulation of matter and gauge fields and atom-number-resolved detection into an updated LGT simulator. With these technical advances, we can monitor the local Gauss's law and then perform the post-selection, which eliminates the processes violating local gauge symmetries. We can measure the order parameter with various system sizes to perform a proper finite-size scaling, which overcomes the finite-size effects and determines the critical point accurately. We can also deterministically prepare the $\ket{\mathbb{Z}_2}$ state by addressing single atoms in a programmed manner. Equipped with these capabilities, we observe the nontrivial thermalization dynamics across the quantum critical regime \cite{Yao:2022qm}.

\begin{figure*}[ht]
	\centering
	\includegraphics[width=0.95\textwidth]{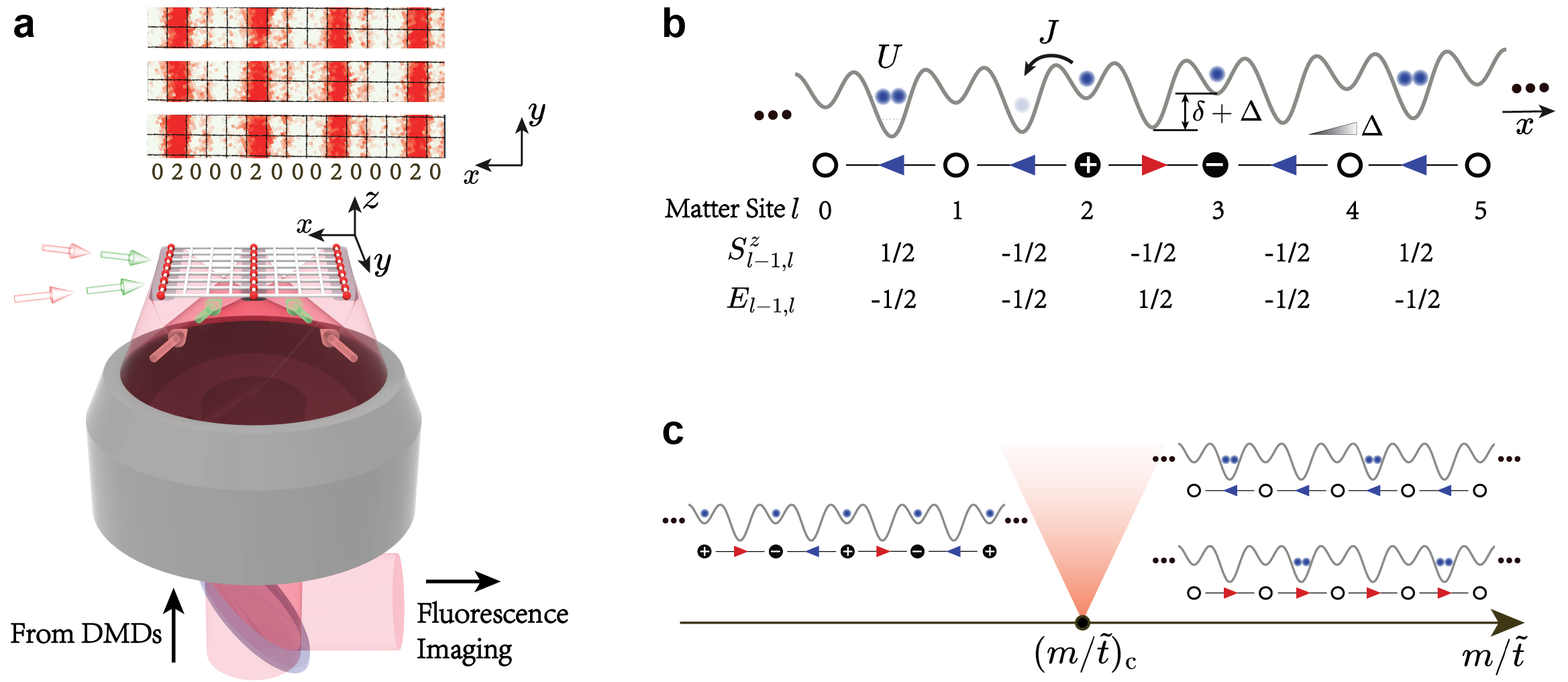}
	\caption{\textbf{Experimental system}. (a) Schematic of the ultracold atom microscope and the prepared $\ket{\mathbb{Z}_2}$ initial state. We combine the optical superlattices and the addressing beam generated by the digital micromirror device (DMD) to prepare the initial $\ket{\mathbb{Z}_{2}}$ state (See Methods). The top shows an exemplary raw-data fluorescence image of the atom distribution of the initial $\ket{\mathbb{Z}_{2}}$ state in a single experimental realization. (b) The physical model with bosons in a one-dimensional optical lattice. Here, $U$ denotes the on-site interaction strength, $J$ denotes the hopping amplitude of bosons, $\delta$ denotes the energy offset between neighboring shallow and deep lattices, and $\Delta$ denotes the linear tilt per site. The open and solid circles with $+$ or $-$ denote physical charge zero, $+1$ or $-1$ at the matter sites, and the arrows denote the electric field. (c) An Ising-type quantum phase transition by tuning $m/\tilde{t}$.
}
	\label{system}
\end{figure*}

Our experimental system is shown in Fig.~\ref{system}(a), which realizes a U(1) LGT using bosons in optical lattices, and the protocol is the same as that reported in the previous work \cite{Yang:2020oo}. We combine a short lattice and a long lattice with twice the lattice spacing to create a one-dimensional superlattice, as shown in Fig.~\ref{system}(b) \cite{Zhang:2022fb}. 
When the deep lattice sites are doubly occupied, the on-site energy is $U-2\delta$, where $U$ is the on-site interaction energy and $\delta$ is the energy offset between deep and shallow lattices. When $U\approx 2\delta$, the on-site energy of a doubly occupied site is nearly degenerate with an empty site, but is off-resonant with a singly occupied state. Thus, we only retain the empty and the doubly occupied states, and they are denoted by $\ket{\downarrow}$ and $\ket{\uparrow}$, respectively. These deep lattice sites are viewed as gauge sites, and the shallow lattice sites are viewed as matter sites. This realizes the LGT written as \cite{Yang:2020oo}
\begin{equation}
\hat{H}=\sum_{l}\left[ \tilde{t} \left(\hat{\psi}_l\hat{S}^{+}_{l,l+1}\hat{\psi}_{l+1}+\text{h.c.}\right)+m\hat{\psi}^\dag_{l}\hat{\psi}_l\right], \label{Hamiltonian}
\end{equation}
where $l$ labels the matter sites, and $\hat{\psi}_l$ are bosonic operators for the matter fields. Spin-$1/2$ operators $\hat{S}_{l,l+1}$ are defined in the deep lattices. Here $m=\delta-U/2$ is the mass of the matter field. $\tilde{t}$ is generated by a second-order hopping process because the long-range direct hoppings are suppressed by applying a tilting potential.

\begin{figure*}[htbp]
	\centering
	\includegraphics[width=0.95\textwidth]{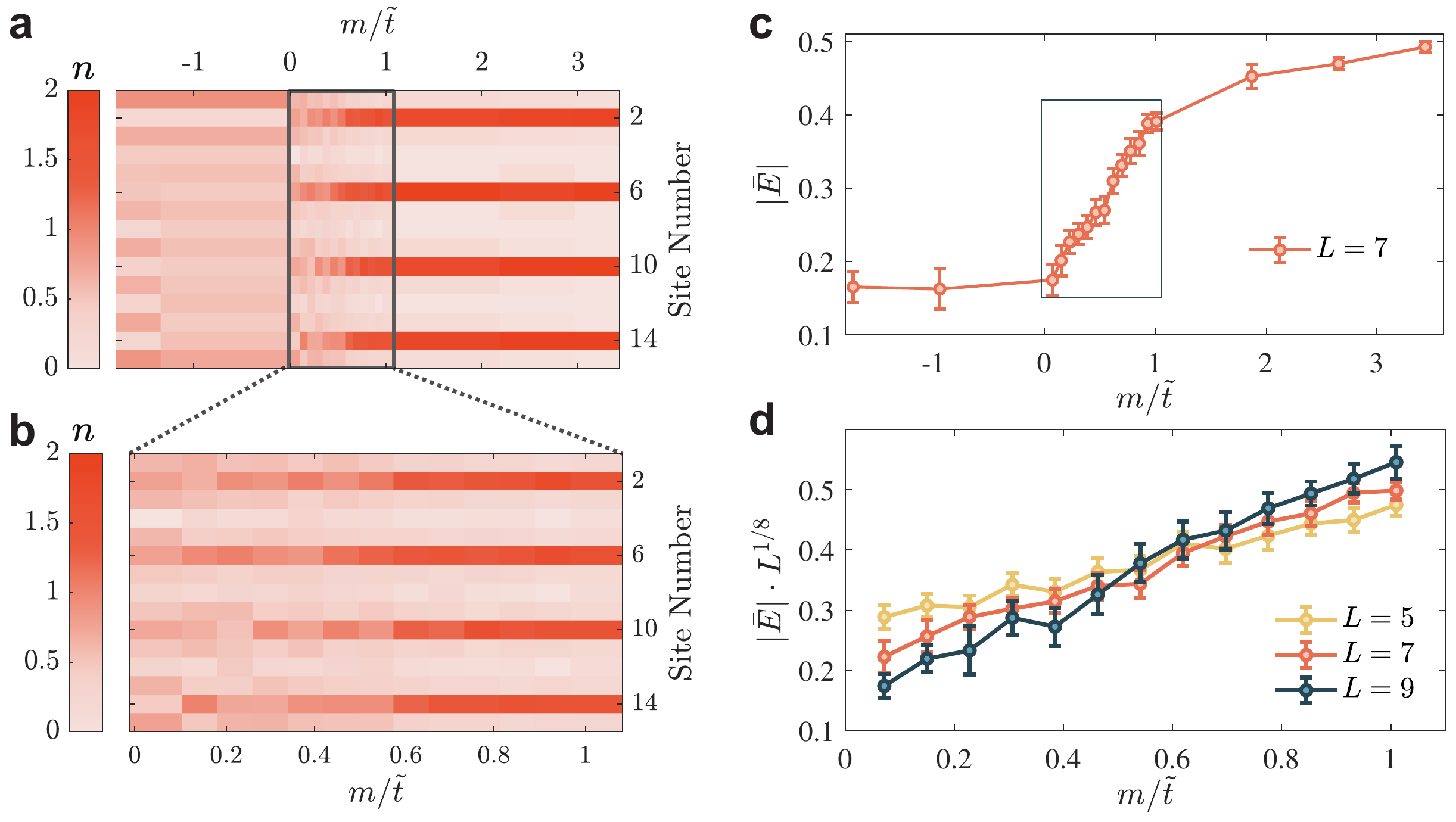}
	\caption{\textbf{Adiabatic ramping and phase transition}. (a-b) Single-site-resolved atom number distribution for a range of $m/\tilde{t}$. (c) The absolute of the spatial averaged electric field $|\bar{E}|$ as a function of $m/\tilde{t}$. (d) $|\bar{E}|L^{1/8}$ as a function of $m/\tilde{t}$ for $L=5,7, 9$. The crossing point of the three curves locates the quantum critical point. Error bars denote the s.e.m.
}
	\label{ramping}
\end{figure*}

This model possesses local gauge symmetries because the Hamiltonian Eq.~\eqref{Hamiltonian} is invariant under the local gauge transformation $\hat{\psi}_l\rightarrow e^{i\theta_l}\hat{\psi}_l$, $\hat{S}^+_{l,l+1}\rightarrow e^{-i\theta_l}\hat{S}^+_{l,l+1}$ and $\hat{S}^+_{l-1,l}\rightarrow e^{-i\theta_l}\hat{S}^+_{l-1,l}$ \cite{Yang:2020oo}. This local gauge symmetry gives rise to a set of local conserved quantities $G_l=S^z_{l-1,l}+S^z_{l,l+1}+n_{l}$. By introducing the electric field $E_{l-1,l}=(-1)^l S^z_{l-1,l}$ and the physical charge $Q_{l}=(-1)^l n_l$, the conservation $G_l=0$ can be written as
\begin{equation}
E_{l,l+1} - E_{l-1,l} =Q_l, \label{Gauss}
\end{equation}
which is exactly the Gauss's law of U(1) LGTs \cite{Yang:2020oo}. A configuration of the electric field and  Gauss's law is also schematically shown in Fig.~\ref{system}(b).

When focusing on the gauge sector with all $G_l=0$, there are two phases in the Hamiltonian Eq.~\eqref{Hamiltonian} as tuning $m/\tilde{t}$. This can be easily seen by looking at two limits with $m/\tilde{t}=\pm \infty$. When $m/\tilde{t}\rightarrow +\infty$, the matter field favors $n_l=0$ and therefore $S^z_{l-1,l}+S^z_{l,l+1}=0$, leading to an antiferromagnetic state $\ket{\mathbb{Z}_2}=\ket{\uparrow\downarrow\uparrow\downarrow\dots}$ or $\ket{\downarrow\uparrow\downarrow\uparrow\dots}$ at gauge sites. This leads to two-fold degenerate ground state. When $m/\tilde{t}\rightarrow -\infty$, the matter field favors $n_l=1$ and therefore $S^z_{l-1,l}=S^z_{l,l+1}=-1/2$. This state does not break the original lattice translational symmetry. Hence the transition is a $Z_2$ symmetry breaking transition of the Ising type, and detailed theoretical calculations predict the critical point at $m/\tilde{t}\approx 0.655$ \cite{Sachdev:2002mi,Fendley:2004cd,Rico:2014tn}. Signatures of two different phases have also been observed in the previous experiment \cite{Yang:2020oo}.

To accurately determine the quantum critical point, we first prepare a $\ket{\mathbb{Z}_2}$ state with high fidelity utilizing the ability of single-site addressing (see Methods) at $\delta=447(1)$ Hz, $U=732(2)$ Hz, and $\tilde{t}=23.5(2)$ Hz, corresponding to $m/\tilde{t}=3.44(7)$ \cite{Weitenberg:2011ss}. This $\ket{\mathbb{Z}_2}$ state is an antiferromagnetic state $\ket{\uparrow\downarrow\uparrow\downarrow\dots}$ in the gauge sites and empty sites in the matter sites. Here we prepare $5 \sim 10$ copies of identical chains along the $x$ direction, as shown in Fig.~\ref{system}(a) [See also Fig. \ref{FigureS01:TimeSequence}(f) in Methods]. Each chain contains $L$ gauge sites and $L+1$ matter sites, with a total of $L+1$ atoms in each chain ($L=5,7,9$ as shown below). Then, we adiabatically ramp $\delta$ to the targeted critical regime around $373$ Hz, with a constant ramping speed $\dot{\delta}=2.1$ Hz/$\mathrm{ms}$. Such change of $\delta$ leaves $U$ and $\tilde{t}$ nearly unaffected, and it changes the value of $m/\tilde{t}$ to the critical regime, as shown in Fig.~\ref{ramping}. Finally, the system is suddenly frozen and the atom number at each site is read out with the single-site-resolved microscope. Note that usually the fluorescence imaging cannot distinguish two atoms from zero atom at the same site \cite{Bakr:2009aq,Sherson:2010sa}. Here, before detection, we split two atoms into two sites of a double well along the $y$ direction if there are two atoms at the same site. This allows us to resolve atom number two from zero at a single site (see Methods). To mitigate the undesired effects from processes beyond the LGT, we post-select our data based on two rules: (i) the total atom number remains the same as that of the initial state, and (ii) the Gauss's law Eq.~\eqref{Gauss} has to be obeyed for all sites. With the adiabatic ramping and the post-selection, we ensure that the ground state of the LGT at the targeted $m/\tilde{t}$ is reached. The results of single-site-resolved measurement of atom number distribution is shown in Fig.~\ref{ramping}(a) for a range of $m/\tilde{t}$.

\begin{figure*}[htbp]
	\centering
	\includegraphics[width=0.95\textwidth]{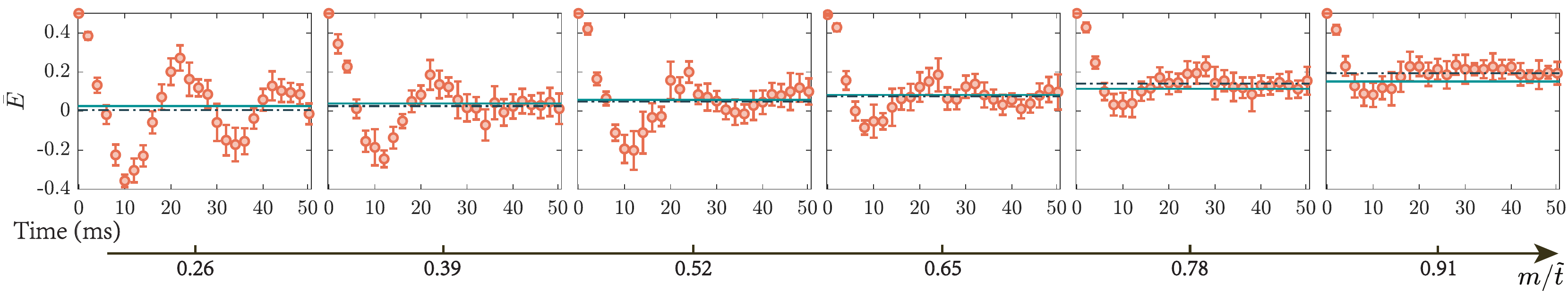}
	\caption{\textbf{Time evolution after a quench}. The real-time dynamics of $\bar{E}$ for different values of $m/\tilde{t}$ (marked at the axes of $m/\tilde{t}$). The dashed lines are the fitted long-time steady values $E_\infty$, and the solid lines are the theoretical thermal values $E_\text{th}$ assuming that the initial state fully thermalizes. Here $L=7$ and each data point is averaged over $10\sim 20$ chains after the post-selection. Error bars denote the s.e.m, and are smaller than the markers if hidden.
}
	\label{quench}
\end{figure*}

\begin{figure}[htbp]
	\centering
	\includegraphics[width=0.48\textwidth]{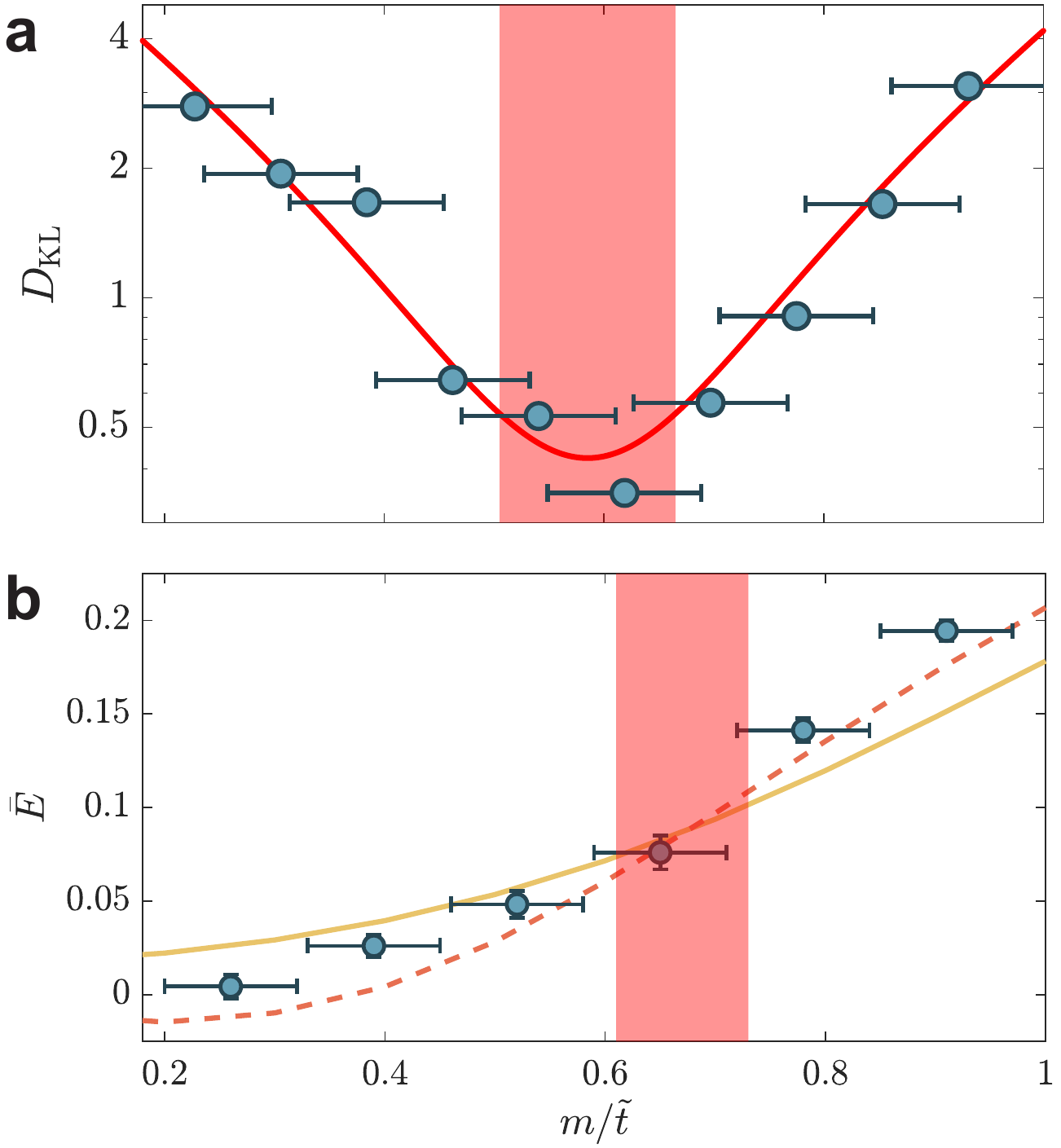}
	\caption{\textbf{Quantum criticality}. (a) The averaged KL divergence $\text{D}_{\text{KL}}$ defined by Eq.~\ref{KLD} quantifies the overlap of three data points for each $m/\tilde{t}$ in Fig.~\ref{ramping}(d). $\text{D}_{\text{KL}}$ is plotted as a function of $m/\tilde{t}$. The peak of $\text{D}_{\text{KL}}$ defines the point where the three curves in Fig.~\ref{ramping}(d) cross each other and locates the quantum critical point. The error bars of the data points are the error of the calibrated $m/\tilde{t}$ (see Methods). The shade region represents the error bar contains both the Gaussian fitting error and the error of the calibrated $m/\tilde{t}$ and its central value is the fitted quantum critical point. (b) The data points are the steady values $E_\infty$ extracted from Fig.~\ref{quench}. The red dashed line is the theoretical steady value with the same system size as the experimental system, and the solid yellow line is the theoretical thermal value $E_\text{th}$ assuming that the initial state obeys the eigenstate thermalization hypothesis. The horizontal error bars denote the errors of the calibrated $m/\tilde{t}$ and the vertical error bars denote the standard deviations of the fitted $E_\infty$. The shade region represents the error of the calibrated $m/\tilde{t}$ and its central value is the point of intersection between theorical thermal value and fitted curve of the steady values $E_\infty$.
}
	\label{transition}
\end{figure}

To locate the critical point, we measure the order parameter $L^{-1}\sum_{l}(-1)^l S^z_{l-1,l}$. This order parameter is the staggered magnetization, and given the definition of the electric field, this order parameter is also the spatial averaged electric field strength, denoted by $\bar{E}$. We plot $|\bar{E}|$ as a function of $m/\tilde{t}$ with $L=7$ in Fig.~\ref{ramping}(c), which exhibits a rapid change when $m/\tilde{t}$ is in the range $\sim [0,1]$. To more accurately determine the critical point, we zoom in this range and perform a finite-size scaling. We plot $|\bar{E}|L^{\beta/\nu}$ for different system size $L=5,7,9$, and for the two-dimensional Ising universality class, the order parameter critical exponent $\beta=1/8$ and the correlation length critical exponent  $\nu=1$ \cite{Francesco:Book}. The crossing point of these curves locates the critical point (see Methods for numerical simulation). Here we indeed observe the crossing of three curves. However, since each data point has an error bar, it is not clear to visualize the exact crossing point. Hence, we consider each point as a normalized Gaussian distribution $p_i(x) = \mathcal{N}(\mu_i, \sigma_i^2)$
% $p_{i}(x)=\exp [-(x-\mu_{i})^2/(2 \sigma_{i}^{2}})]/ (\sqrt{2 \pi} \sigma_{i})$,
where $x =|\bar{E}|L^{1/8}$, $\mu_{i}$ and $\sigma_i$ are respectively the value and the error bar of each data point. We calculate the averaged Kullback--Leibler (KL) divergence of the three Gaussian distributions at each $m/\tilde{t}$, which quantifies the degree of overlap between three Gaussian distributions. The averaged KL divergence is defined as \cite{Sgarro:1981id}
\begin{equation}
\text{D}_{\text{KL}}=\frac{1}{6} \sum_{i,j=1}^{3}\int_{-\infty}^{+\infty} dx p_{i}(x)\log\left(\frac{p_i(x)}{p_j(x)}\right). \label{KLD}
\end{equation}
We plot the $\text{D}_{\text{KL}}$ as a function of $m/\tilde{t}$ in Fig.~\ref{transition}(a), and fitting the data yields a peak position at $m/\tilde{t}=0.59(8)$. This value is consistent with theoretical predictions \cite{Sachdev:2002mi,Fendley:2004cd,Rico:2014tn}. Given the energy scale of our $\tilde{t}$, the deviation from the theoretical critical value is less than $2$ Hz.

We then consider the quench dynamics. The same $\ket{\mathbb{Z}_2}$ state is prepared as the initial state. Now, instead of adiabatic ramping, we suddenly change the value of $m/\tilde{t}$ to the targeted value nearby the critical point, with a typical time scale of $1~\mathrm{ms}$. Here we fix $\tilde{t}=34.1(3)$ Hz and $U=676(2)$ Hz. The value of $\tilde{t}$ is larger than what we used for the adiabatic ramping process. This is because we find that even a weak spatial inhomogeneity can cause noticeable variations of $m$ from site to site, which can affect the time dynamics. A larger $\tilde{t}$ helps to reduce this variation in terms of $m/\tilde{t}$ and can significantly suppress the effects of the inhomogeneity. We record the spatial atom number distribution at different times during the real-time dynamics using a single-site-resolved microscope. We also apply the same post-selection to ensure that the dynamics is governed by the LGT. The measurements of time dynamics are shown in Fig.~\ref{quench} for different values of $m/\tilde{t}$.

We fit the experimental data in Fig.~\ref{quench} with a damped sinusoidal function $Ae^{-t/\tau}\sin(\omega t)+E_\infty$, where $A$, $\tau$, $\omega$, and $E_\infty$ are all fitting parameters. We obtain $E_\infty$ as the long-time steady value, as illustrated by the dashed lines in Fig.~\ref{quench}. Meanwhile, we can also theoretically extract the thermalization values $E_\text{th}$, provided that the system obeys the eigenstate thermalization hypothesis and the initial $\ket{\mathbb{Z}_2}$ state thermalizes \cite{Deutsch:1991qs,Srednicki:1994ca,Rigol:2008ta,DAlessio:2016fq}. $E_\text{th}$ is obtained by calculating $\text{Tr}\left[\rho(T)L^{-1}\sum_{l}(-1)^l S^z_{l-1,l}\right]$. Here $\rho(T)$ is an equilibrium density matrix of the LGT system, with the temperature $T$ determined by matching energy $\text{Tr}[\rho(T)\hat{H}]=\bra{\mathbb{Z}_2}\hat{H}\ket{\mathbb{Z}_2}$ (see Methods). Previous work has predicted that for the PXP model, $E_\infty$ matches $E_\text{th}$ and the initial $\ket{\mathbb{Z}_2}$ state thermalizes only in the quantum critical regime \cite{Yao:2022qm}. These two values depart from each other away from the critical point $(m/\tilde{t})_\text{c}$. When $m/\tilde{t}>(m/\tilde{t})_\text{c}$, the ground state is two-fold degenerate and the $\ket{\mathbb{Z}_2}$ state has large overlap with the ground state, and the ground state is always not thermalized. When $m/\tilde{t}<(m/\tilde{t})_\text{c}$, especially around $m/\tilde{t}=0$, it is known that the PXP model hosts a set of many-body scar states as the system's eigenstates \cite{Bernien:2017pm,Turner:2018we,Turner:2018qs}. The $\ket{\mathbb{Z}_2}$ state also has large overlap with the scar states, preventing it from thermalization \cite{Bernien:2017pm,Turner:2018we,Turner:2018qs}. The PXP model is equivalent to this LGT model under the local gauge constraints of $G_l=0$ for all $l$, and therefore, the discussion also applies to this LGT model \cite{Surace:2020lg,ChengTA}. In Fig.~\ref{transition}(b), we compare $E_\infty$ with $E_\text{th}$, and our measurements agree with the prediction that  $E_\infty \approx E_\text{th}$ only in the quantum critical regime \cite{Yao:2022qm}.

In summary, we have performed a single-site-resolved quantitative experimental study of the quantum criticality in the U(1) LGT realized with bosons in optical lattices. Our study combines both the equilibrium property and the thermalization dynamics. Our system size is up to $\sim 19$ lattice sites, and our quantitative results agree well with the numerical results of exact diagonalization. This agreement benchmarks the validity of our ultracold atom quantum simulator quantitatively, and demonstrates this simulator as a powerful platform to study non-equilibrium dynamics of the gauge theory. In the near future, when our system size is enlarged several times, it will be beyond the capability of exact diagonalization. The experimental control and detection capability developed in this work can be used to study other interesting dynamical phenomena in this system, such as string breaking \cite{Banerjee:2012aq,Hebenstreit:2013rt}, dynamical transition between quantum phases \cite{Huang:2019dq,Zache:2019dt}, the false vacuum decay, and the confinement-deconfinement transition \cite{Surace:2020lg,ChengTA,Hauke}. The current scheme of implementing the LGT can also be extended to higher dimensions \cite{Ott:2021sc}.

\textbf{Acknowledgement} 
    We thank Meng-Da Li, Qian Xie, Bo Xiao, Hui Sun, Wan Lin, An Luo for their help on building the experimental setup. This work was supported by NNSFC grant 12125409, Innovation Program for Quantum Science and Technology 2021ZD0302000.
    
%\textbf{Competing interests:}
%    The authors declare no competing interests.

%\textbf{Data availability:}
%    Data used in this work is available on reasonable request.

%\textbf{Code availability:}
%    Code used in this work is available on reasonable request.

%%%%%% Supplementary materials
\newpage
\onecolumngrid
\vspace*{0.5cm}
\newpage
\begin{center}
    \textbf{METHODS AND SUPPLEMENTARY MATERIALS}
\end{center}
\vspace*{0.5cm}
\twocolumngrid
\setcounter{equation}{0}
\setcounter{figure}{0}
\makeatletter
\makeatother
\renewcommand{\theequation}{S\arabic{equation}}
\renewcommand{\thefigure}{S\arabic{figure}}
\renewcommand{\thetable}{S\arabic{table}}

\section{Experimental Procedure}

\begin{figure*}[htb]
    \centering     %
    \includegraphics[width=150mm]{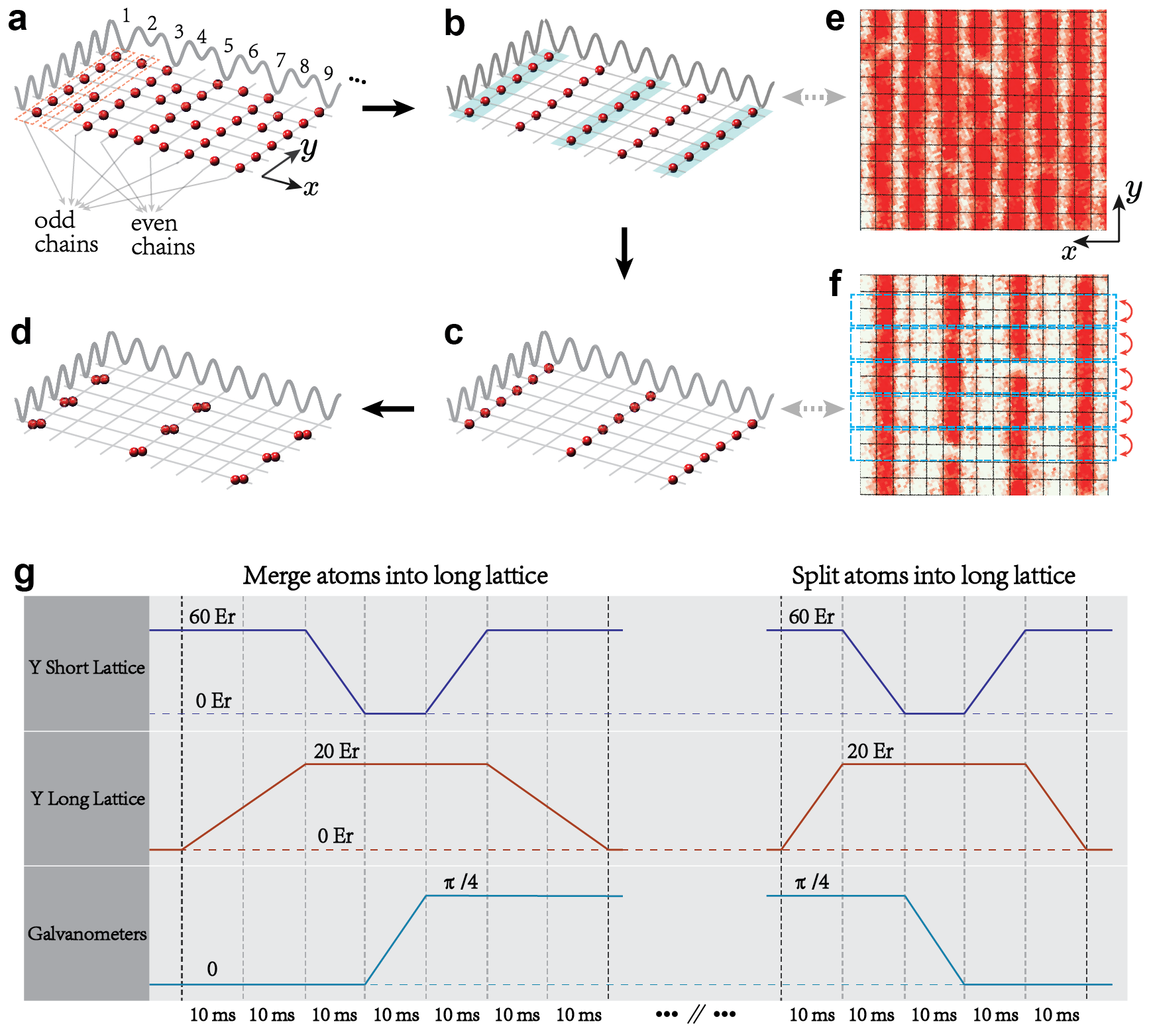}
    \caption{\textbf{Initial state preparation and atom number detection}.
    (a) An illustration of a typical snapshot of the atom distribution of the prepared state after staggered-immersion cooling. (b) The unity filling atom chains are obtained by removing the atoms in the ``even chains'' of (a) via site-dependent addressing. Atoms in blue shaded chains are then addressed by DMD light. (c) An illustration of the state after removing the unaddressed atoms in (b). (d) The prepared copies of $\ket{\mathbb{Z}_2}$ state along the $x$ direction after we merge every two atoms onto a single site along the $y$ direction. (e) and (f) are two typical fluorescence images of the states illustrated in (b) and (c), respectively. (g) The time sequences of the ``short lattice" potential, the ``long lattice" potential and the relative phase $\theta_y$ controlled by galvanometers. These two processes are used to merge two atoms into a single site and split atoms in one site into a double well respectively.
    }
    \label{FigureS01:TimeSequence}
\end{figure*}

\noindent\textbf{Initial state preparation.}
Our experiment begins by preparing a two-dimensional (2D) Bose-Einstein condensate of $^{87}$Rb atoms in the $\ket{F=1,m_\mathrm{F}=-1}$ state, which has been described in our previous work \cite{Xiao:2020gt,Zhang:2022fb}. We then employ the recently demonstrated staggered-immersion cooling method to prepare a near-unity filling Mott insulator state \cite{Yang:2020ca,Zhang:2022fb}. As a result, we obtain a series of one-dimensional (1D) Mott insulators with a filling factor of $99(1)\%$ prepared in the ``odd chains'' along the $y$ direction, as schematically shown in Fig.~\ref{FigureS01:TimeSequence}(a). Then, we remove the atoms in the ``even chains'' via site-dependent addressing \cite{Yang:2017sd}, leaving a system with 1D Mott insulators only in the ``odd chains", as shown in Fig.~\ref{FigureS01:TimeSequence}(b). The next step is addressing the Mott-insulator chains in an alternating fashion, using a beam of wavelength $787.55~\mathrm{nm}$ reflected from a DMD \cite{Weitenberg:2011ss}. For example, only the blue shaded chains in~\ref{FigureS01:TimeSequence}(b) are addressed. After that, we remove the unaddressed chains to prepares a series of Mott-insulator chains separated by $4 a_{\mathrm{short}}$, as shown in Fig.~\ref{FigureS01:TimeSequence}(c). The last step is to turn on the superlattice potential (not shown in Fig.~\ref{FigureS01:TimeSequence}) in the $y$ direction to merge every two atoms into one site \cite{Yang:2020ca}. We end up with copies of initial state $\ket{\mathbb{Z}_{2}} =\ket{020002000\dots }$ along the $x$ direction, as shown in Fig.~\ref{FigureS01:TimeSequence}(d). In terms of gauge field degrees of freedom, this state reads $\ket{\mathbb{Z}_{2}}=\ket{\uparrow\downarrow\uparrow\downarrow\dots}$.

\noindent\textbf{Atom number detection.}
To obtain the single-site-resolved atom occupancy of the final state, we split the atoms at each site into a double well along the $y$ direction using the time sequence shown in Fig.~\ref{FigureS01:TimeSequence}(g). Assuming that there are at most two atoms at the same site, this splitting process has three possible outcomes: (i) $\ket{20} \to \ket{11}$; (ii) $\ket{10} \to \ket{01}$($\ket{10}$); and (iii) $\ket{00} \to \ket{00}$. Finally, we perform a fluorescence imaging to record the atom parity distribution that can be used to reconstruct the atom number distribution before the splitting process \cite{Zhang:2022fb}.

\begin{figure}[htb]
    \centering
    \includegraphics[width=75mm]{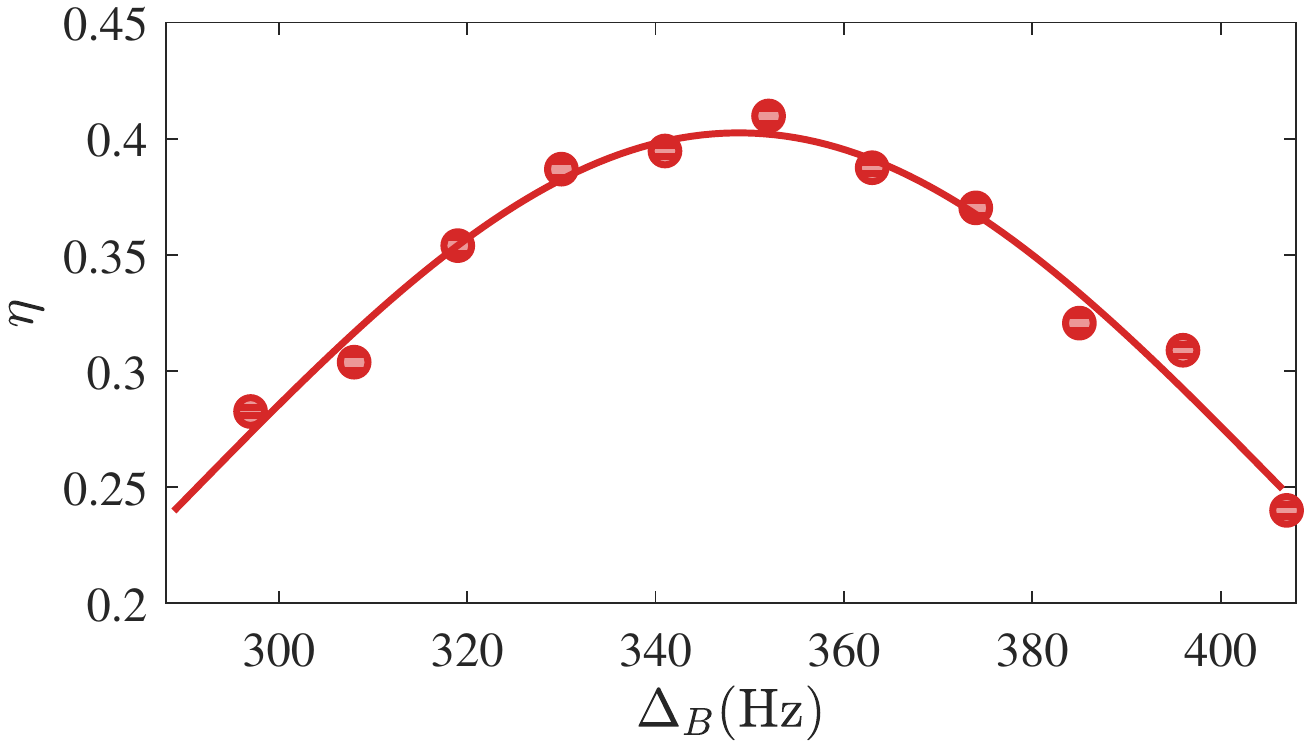}
    \caption{\textbf{Calibrations of staggered potential $\delta$}.
    The data points represent the population fraction of the atoms that tunnel to adjacent sites after applying the magnetic field gradient. The peak position of the Gaussian fit corresponds to $\delta \approx \Delta_{\mathrm{B}}$ with $\delta = 349(1)$ Hz, and the error bar is the standard deviation of the fitted $\delta$. Error bars denote the s.e.m and can be smaller than the size of the markers.
    }
    \label{FigureS02:CalibrationsD}
\end{figure}

\section{Calibrations}
\noindent \textbf{Calibration of Hubbard parameters}.
The lattice depths of the ``short lattice" and the ``long lattice'' along the $x$ and $y$ directions are calibrated using lattice modulation spectroscopy, the details can be found in our previous work \cite{Zhang:2022fb}. The Hubbard parameters, tunneling strength $J$ and on-site interaction strength $U$, are also calibrated with the same method described in our previous work \cite{Zhang:2022fb}.

\noindent \textbf{Calibration of staggered potential $\delta$}.
In the experiment, the staggered potential $\delta$ is controlled by the depth of the ``long lattice'' along the $x$ direction. To calibrate $\delta$, we start with the initial state shown in Fig.~\ref{FigureS01:TimeSequence}(c). Then, we ramp up the ``long lattice" along the $x$ direction with $\theta_{x} = \pi / 4$ to create the same staggered structure as that in Fig.~\ref{system}(b). Next, a magnetic field gradient is applied to create an energy bias $\Delta_{\mathrm{B}}$ between two adjacent lattice sites along the $x$ direction. After that, we ramp down the ``short lattice'' along the $x$ direction from $60E_{\mathrm{r}}$ to $12E_{\mathrm{r}}$ in $300~\mu{\mathrm{s}}$ while keeping the ``short lattice'' along the $y$ direction at $60E_{\mathrm{r}}$ and let the system evolve for $8~{\mathrm{ms}}$. Finally, we record the positions of the atoms using single-site-resolved fluorescence imaging. We measure the population fraction $\eta$ of the atoms that tunnel to its adjacent site as a function of the energy bias $\Delta_{\mathrm{B}}$, as shown in Fig.~\ref{FigureS02:CalibrationsD}. Since $\delta \gg J$, this function $\eta$ should be peaked at $\delta \approx \Delta_{\mathrm{B}}$, where the atoms undergo nearly resonant Rabi oscillation between adjacent sites. By fitting the data with a Gaussian function, we can extract the corresponding staggered potential $\delta$. From the calibrated staggered potential $\delta$ and the interaction strength $U$, we can determine the mass using $m=\delta - U/2$.

\begin{figure}[tb]
    \centering
    \includegraphics[width=75mm]{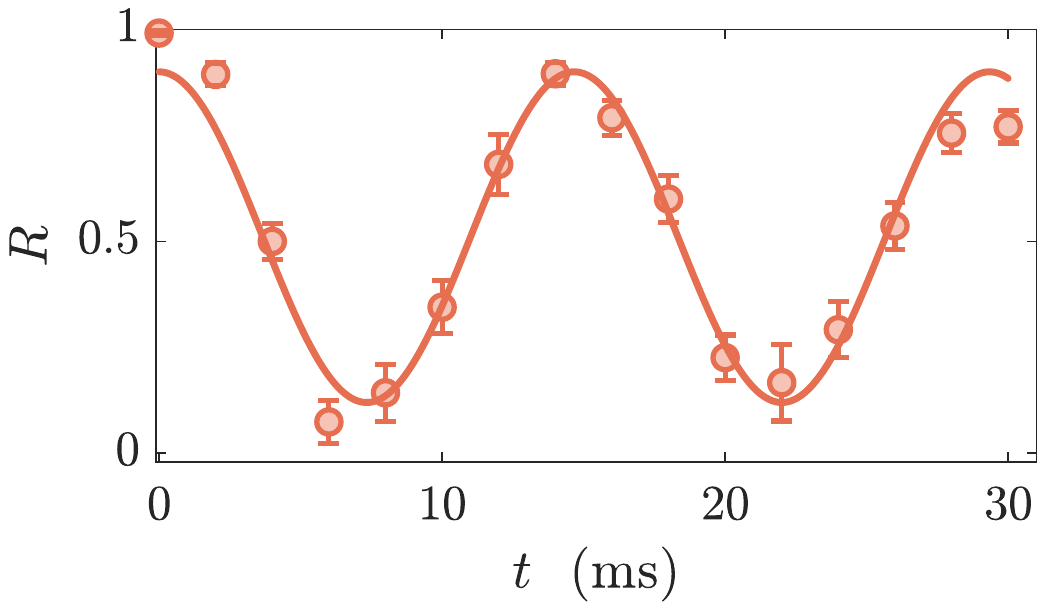}
    \caption{\textbf{Calibration of $\tilde{t}$}.
    We observe coherent oscillation between $\ket{\varphi_{0}} =\ket{020}$ and $\ket{\varphi_{1}} =\ket{101}$ for three sites with two atoms. Here $R = n_{020} / (n_{020} + n_{101})$ is the population fraction of the $\ket{020}$ state. The solid curve is a sinusoidal fitting result, which gives an oscillation frequency of $34.1(3)~\mathrm{Hz}$. The error bar of the oscillation frequency is the s.d of the fitted parameter and the error bars of the data points denote s.e.m.
    }
    \label{FigureS03:PXP}
\end{figure}

\noindent \textbf{Calibration of $\tilde{t}$}.
To calibrate the pair hopping strength $\tilde{t}$, we prepare multiple isolated systems of three sites with two atoms, and we start with the initial state $\ket{\phi_{0}} = \ket{020}$. Then we quench the parameters $J$ and $U$ of the system to the targeted values while keeping $m/\tilde{t}=0$. After evolving the system for a time $t$, we post-select two states $\ket{\varphi_{0}} =\ket{020}$ and $\ket{\varphi_{1}} =\ket{101}$. Let $n_{020}$ and $n_{101}$ denote the number of occurrences that we detect $\ket{\varphi_{0}}$ and $\ket{\varphi_{1}}$ respectively. We then plot the population fraction $R = n_{020}/(n_{020}+n_{101})$ in Fig.~\ref{FigureS03:PXP} as a function of the evolution time. The pair hopping strength $\tilde{t}$ can be read out from the oscillation frequency of $R$ fitted by a sinusoidal function.

\begin{figure}[tb]
    \centering     
    \includegraphics[width=75mm]{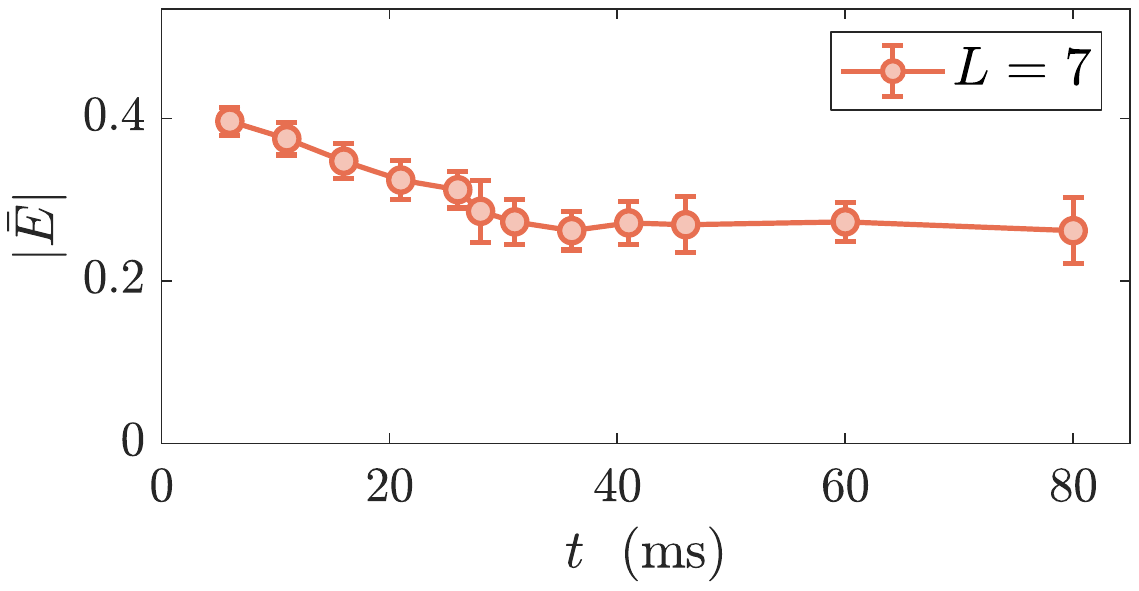}
    \caption{\textbf{Checking adiabatic condition.} Plot of the $\left | \bar{E} \right|$ of the final state against the ramping time starting from the $\ket{\mathbb{Z}_2}$ initial state. The system size is $L=7$, and the staggered potential $\delta$ is tuned from $\delta=447(1)$ Hz to $\delta=372(1)$ Hz during the ramping process. The error bars denote s.e.m.}
    \label{FigureS04:AdiRamping}
\end{figure}

\noindent\textbf{Adiabatic condition}.
To determine the adiabatic condition of the ramping process, we prepare the initial $\ket{\mathbb{Z}_{2}}$ state and ramp $\delta$ from $=447(1)$ Hz [$m/\tilde{t}=3.44(7)$] to $372(1)$ Hz [$m/\tilde{t}=0.62(7)$] with different ramping speeds. Then we plot the $\left| \bar{E} \right|$ of the final state as a function of the ramping time. As one can see from Fig.~\ref{FigureS04:AdiRamping}, when the ramping time is longer than $30~\mathrm{ms}$, the observable $\left | \bar{E} \right|$ of the final state barely changes with the ramping time. This shows that the adiabatic condition is satisfied in this regime. In the experiment, we ramp $\delta$ from $447(1)$ Hz to $372(1)$ Hz in $36~\mathrm{ms}$ at a speed of $\dot{\delta}=2.1$ Hz/$\mathrm{ms}$ to fulfill the adiabatic condition.

\section{Numerical Simulation}

\noindent\textbf{Finite-size scaling.} According to the theory of finite-size scaling (FSS) \cite{Cardy:Book_FSS,Binder:Book}, the absolute value of the order parameter scales with system size $L$ as
\begin{equation} \label{eq:ffs}
	|\bar{E}| = L^{-\beta/\nu} f(L^{1/\nu} \epsilon).
\end{equation}
Here, $|\bar{E}| = \langle |\hat{\bar{E}}| \rangle =  \langle | L^{-1} \sum_{l} (-1)^{l} \hat{S}_{l-1, l}^{z}| \rangle$ where the average $\langle \cdots \rangle$ is the quantum average over the ground state. $\beta$ is the order parameter critical exponent, $\nu$ is the correlation length critical exponent. $\epsilon = (g - g_{\text{c}})/g_{\text{c}}$ is the reduced ``temperature" with $g=m/\tilde{t}$ being the control parameter and $g_{\text{c}}$ being the critical point, and $f(x)$ is the scaling function. Therefore, the critical point $g_{\text{c}}$ can be determined as the crossing point of different curves of $|\bar{E}| L^{\beta/\nu}$ plotted against $m/\tilde{t}$.

\begin{figure}[t]
	\centering
	\includegraphics[width=0.48\textwidth]{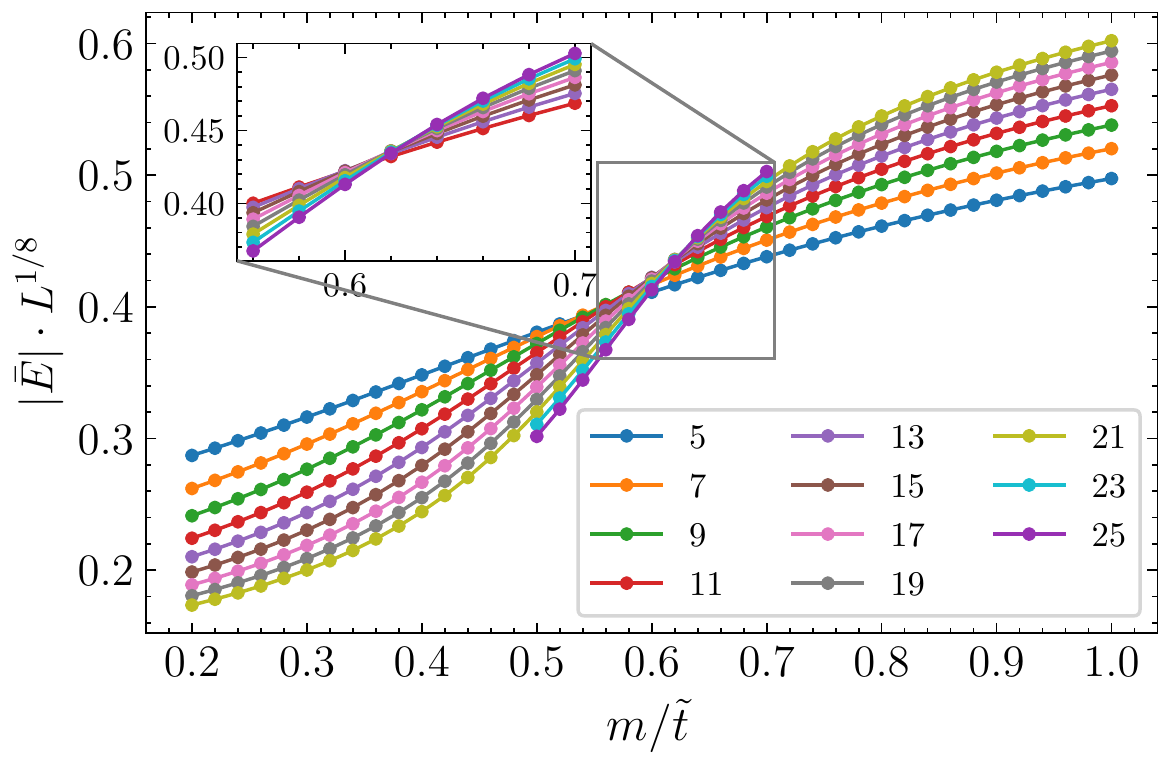}
	\caption{Finite-size scaling plot of $|\bar{E}|L^{\beta/\nu}$ against $m/\tilde{t}$ with system sizes ranging from $5$ to $25$ (shown in the legend). The inset is the same plot for systems sizes from $11$ to $25$. The exact values of two-dimensional Ising critical exponents $\beta=1/8$ and $\nu=1$ are used in the plot \cite{Francesco:Book}.}
	\label{fig:Figure_S05}
\end{figure}

Gauss's law, $S^{z}_{l-1, l} +S^{z}_{l, l+1} + n_l = 0$ in the gauge sector $G_{l}=0$, means that the matter field configuration is completely determined by that of gauge sites. Therefore, the U(1) LGT \eqref{Hamiltonian} can be written in terms of the gauge fields only,
\begin{equation} \label{eq:PXPm}
	\hat{H} = \tilde{t} \sum_{l} \hat{X}_{l, l+1} - m \sum_{l} \hat{Z}_{l, l+1}
\end{equation}
where $\hat{X}_{l, l+1} = 2 \hat{S}^{x}_{l, l+1}$ and $\hat{Z}_{l, l+1} = 2 \hat{S}^{z}_{l, l+1}$ are the standard Pauli operators. In the following, we shall employ this gauge field only Hamiltonian to perform exact diagonalization study under the open boundary condition to calculate $|\bar{E}|$. And this Hamiltonian is precisely the PXP model with an additional external field. The resulting finite-size scaling plot of the scaling variable $|\bar{E}|$ is shown in Fig.~\ref{fig:Figure_S05}.

\begin{figure}[t]
	\centering
	\includegraphics[width=0.9\linewidth]{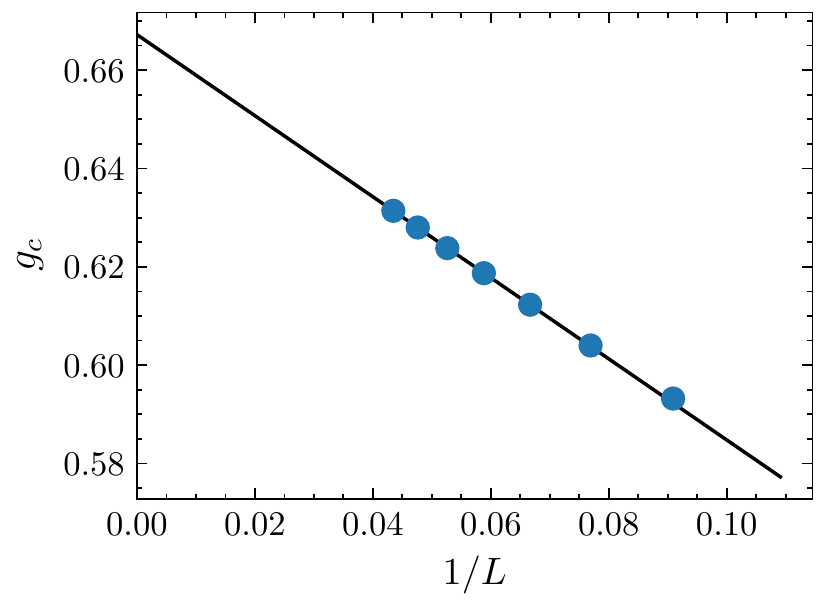}
	\caption{Plot of the crossing point (blue dot) of system size $L$ and $L+2$ against $1/L$ using the scaling variable $|\bar{E}|$. The black line is a linear fit of the first five data points. The intercept of the linear fit gives us the estimated critical point $g_{\text{c}} = 0.67$. }
	\label{fig:Figure_S06}
\end{figure}

Due to corrections to scaling, different curves will not cross at the single critical point. We fit the curves in the inset of Fig.~\ref{fig:Figure_S05} and then calculate the crossing point between the curve of system size $L$ and the curve of system size $L+2$, which yields $g_\text{c}(L)$. Extrapolating $g_{\text{c}}(L)$ to infinite size determines the critical point. In this way, our estimate of the critical point is $g_{\text{c}} = (m/\tilde{t})_{\text{c}} = 0.67$, as shown in Fig.~\ref{fig:Figure_S06}. Besides the order parameter, variables such as $m_{\text{s}}^{2} \equiv \langle \hat{m}_{\text{s}}^{2} \rangle, m_{\text{s}}^{4} \equiv \langle \hat{m}_{\text{s}}^{4} \rangle$ (here $\hat{m}_{\text{s}} \equiv \hat{\bar{E}}$), and the Binder cumulant $R = 1 - m_{\text{s}}^{4}/3m_{\text{s}}^{2}$ can also be used for FSS analysis \cite{Binder:Book}. Following the same procedure described above, the estimated critical point is $g_{\text{c}} = 0.66$ using $m_{\text{s}}^{2}$ and $g_{\text{c}} = 0.68$ using $R$ for FFS analysis. Taking into these slightly different values of $g_{\text{c}}$ into account, our estimated critical point is thus $g_{\text{c}} = 0.67(1)$, in agreement with the more accurate estimate of $g_{\text{c}} = 0.655(3)$ \cite{Rico:2014tn}. In summary, our numerical simulation demonstrates the applicability of measuring $|\bar{E}|$ with different system sizes to accurately determine the critical point.

\noindent \textbf{Computation of the thermalization values and the steady values}.
For a quantum system after a quench, a physical observable $\hat{O}$ is expected to fluctuate around a steady value $O_{\infty}$ after the relaxation time $t_{\text{re}}$. On the other hand, its thermal value $O_{\text{th}}$ is defined to be micro-canonical ensemble average of $\hat{O}$ at the energy of the initial state. Due to the equivalence of statistical ensembles, the thermal value can also be calculated using the canonical ensemble at some temperature $T$
\begin{equation} \label{eq:O_th}
	O_{\text{th}} = \operatorname{Tr} \left( \hat{\rho}(T) \, \hat{O} \right)  \, , \quad \hat{\rho}(T) = \frac{\text{e}^{-\hat{H}/T}}{\operatorname{Tr} e^{-\hat{H}/T} } \, .
\end{equation}
The temperature $T$ is determined by equating the ensemble averaged energy with the energy of the initial state
\begin{equation} \label{eq:T}
	\operatorname{Tr} \left( \hat{\rho}(T) \, \hat{H} \right) = \bra{\psi_{0}}\hat{H}\ket{\psi_{0}} \, ,
\end{equation}
where $\ket{\psi_{0}}$ is the initial state. As explained in the previous section, we can employ the gauge fields only Hamiltonian Eq.~\eqref{eq:PXPm} to perform the calculation. We choose the staggered magnetization, i.e. , the spatial averaged electric field strength $\bar{E}$, as our observable $O$, and proceed to use exact diagonalization to calculate its thermal value according to Eqs.~\eqref{eq:O_th} and \eqref{eq:T}. The result is shown as the solid yellow line in Fig.~\ref{transition}(b).

\begin{figure}[t]
	\centering
	\includegraphics[width=0.48\textwidth]{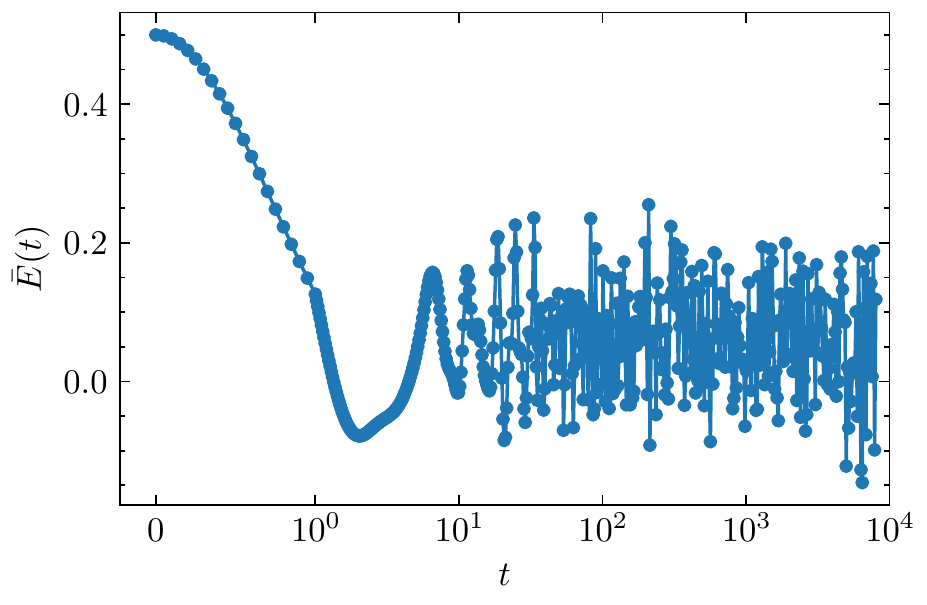}
	\caption{Time evolution of $\bar{E}$ after a quench from the $\ket{\mathbb{Z}_{2}}$ state for system size $L=7$ and parameters $m/\tilde{t}=0.6$. To guide the eye, the linear scale is used for time in $[0, 1]$ and the logarithmic scale is used for longer times. Here, the dimensionless evolution time $t$ is in unit of $\hbar/\tilde{t}$.}
	\label{fig:Figure_S07}
\end{figure}

We then move to determine the relaxation time $t_{\text{re}}$ and the steady value. To this end, starting from the $\ket{\mathbb{Z}_2}$ state, we evolve the system for a long time. A typical case for $L=7$ and $m/\tilde{t}=0.6$ is shown in Fig.~\ref{fig:Figure_S07}. From this plot, we infer that the relaxation time of $\bar{E}$ is less than $10^{2}$ (we use dimensionless time in unit of $\hbar/\tilde{t}$) for this parameter setting. Repeating this process for other values of $m/\tilde{t}$ from $0.2$ to $1.0$, we obtain a safe estimate of the upper bound of the relaxation time of $\bar{E}$ as $t_{\text{re}} < 10^{3}$. By definition, the steady value $E_{\infty}$ can be evaluated as follows
\begin{equation}
	E_{\infty} = \lim_{t_{0} \rightarrow \infty} \dfrac{1}{t_{0}} \int_{t_{0}}^{2 t_{0}} \langle \psi(t) | \hat{\bar{E}} | \psi(t) \rangle \, dt \, ,
\end{equation}
where $|\psi(t)\rangle$ is the wave function at time $t$. In numerical calculation, we choose $t_{0} > t_{\text{th}}$ and average over around $t_{0}$ random sampled time evolution data points in time range $[t_{0}, 2t_{0}]$ to extract $E_{\infty}$. We also increase the value of $t_0$ several times in doing the above average to confirm convergence. We plot thus obtained steady values $E_{\infty}$ as the red dashed line in Fig.~\ref{transition}(b).

\end{document}